\documentclass[12pt]{article}

\usepackage{scicite}
\usepackage{ulem}
\usepackage{times}

\usepackage{graphicx}

\usepackage{color}  
\usepackage{soul}   

\newenvironment{sciabstract}{%
\begin{quote} \bf}
{\end{quote}}

\title{Photoionization time delays probe electron correlations}

\author
{Mingxuan Li,$^{1}$ Huiyong Wang,$^{1}$ Rezvan Tahouri,$^{2}$ Robin Weissenbilder,$^{2}$\\ Jialong Li,$^{1}$  
Wentao Wang,$^{1}$ Jiaao Cai,$^{1}$ 
Xiaochun Hong,$^{1}$ Xiaosen Shi,$^{1}$\\ Liang-Wen Pi,$^{3}$ David Busto,$^{2}$  Mathieu Gisselbrecht,$^{2}$ Kiyoshi Ueda,$^{1,4}$\\ Philipp V. Demekhin,$^{5}$ Anne L'Huillier,$^{2}$ Jan Marcus Dahlström,$^{2*}$\\ Eva Lindroth,$^{6*}$ Dajun Ding,$^{1*}$ Sizuo Luo$^{1*}$
\\
\normalsize{$^{1}$Institute of Atomic and Molecular Physics, Jilin University,}\\
\normalsize{Qianjin Street 2699, Changchun, 130012, China}\\
\normalsize{$^{2}$Department of Physics, Lund University, Box 118, Lund, 221 00 , Sweden}\\
\normalsize{$^{3}$Center for Attosecond Science and Technology, }\\
\normalsize{Xi’an Institute of Optics and Precision Mechanics of the Chinese}\\
\normalsize{Academy of Sciences, Xi’an, 710119, China}\\
\normalsize{$^{4}$Department of Chemistry, Tohoku University, Sendai, 980-8578, Japan}\\
\normalsize{$^{5}$Institut für Physik und CINSaT, Universität Kassel,}\\
\normalsize{Heinrich-Plett-Straße 40, Kassel, D-34132, Germany}\\
\normalsize{$^{6}$Department of Physics, Stockholm University,}\\
\normalsize{AlbaNova University Center, SE-106 91 Stockholm, Sweden}\\
\normalsize{$^\ast$Corresponding Author. E-mail: marcus.dahlstrom@matfys.lth.se (M.D.);} \\
\normalsize{$^\ast$Corresponding Author. E-mail: eva.lindroth@fysik.su.se (E.L.);} \\
\normalsize{$^\ast$Corresponding Author. E-mail: dajund@jlu.edu.cn (D.D.)};\\
\normalsize{$^\ast$Corresponding Author. E-mail: luosz@jlu.edu.cn (S.L.).} 
}

\date{}

\begin{document} 


\baselineskip24pt


\maketitle


\begin{sciabstract}
  The photoelectric effect, explained by Einstein in 1905, is often regarded as a one-electron phenomenon. However, in multi-electron systems, the interaction of the escaping electron with other electrons, referred to as electron correlation, plays an important role. For example, electron correlations in photoionization of the outer $s$-subshells of rare gas atoms lead to a substantial minimum in the ionization probability, which was theoretically predicted in 1972 and experimentally confirmed using synchrotron radiation. However, recent attosecond photoionization time delay measurements in argon strongly disagree with theory, thus raising questions on the nature of electron correlations leading to this minimum. In this work, combining high-spectral resolution attosecond interferometry experiments and novel theoretical calculations allows us to identify the most essential electron correlations affecting the photoemission. The measurement of time delays gives unprecedented insight into the photoionization process, unraveling details of the atomic potential experienced by the escaping electron and capturing its dynamics.  
\end{sciabstract}

\paragraph*{Introduction}

Electron correlations arise due to the mutual repulsion between electrons in matter and determine essential properties of a wide range of quantum systems, from atoms and molecules to surfaces and bulk solid materials~\cite{slaterPR1951simplification,lowdinPR1955quantumCorrelation,dagotto1994RMPcorrelated}. A textbook example of the effect of electron correlations is the {\it Amusia-Cooper minimum} \cite{Amusia_1972_PL_s_subshell,mobusPRA19933Smeasurements,amusia2012handbook} in the photoionization probability of the $s$-valence subshells of rare gas atoms. The {\it Cooper minimum} in the photoionization of the $p$-valence subshells of rare gas atoms \cite{Cooper1962,samson_2002_JESRP_CS3p} is a one-electron effect induced by a change of sign of the dominant ionization channel. The Amusia-Cooper minimum originates from the coupling between the n$s$ and n$p$ ionization channels, which leads to a replica of the n$p$ Cooper minimum, shifted in energy, in the n$s$ photoionization probability. This effect was predicted in a seminal article by Amusia et al. in 1972 and is called the Amusia-Cooper minimum (ACM) in the present work to emphasize the different natures of the two minima. The theoretical prediction by Amusia and coworkers was beautifully verified by experimental measurements of the photoionization cross-sections of rare gases using synchrotron radiation \cite{mobusPRA19933Smeasurements}.

Attosecond spectroscopy provides a new insight into light-matter interaction by giving access to electron motion in the time domain. It has, for example, allowed the determination of photoionization time delays -- the time difference between an electron propagating out of the atomic potential and a freely propagating electron \cite{DahlstrJPB2012Review,PazourakReview}. This concept of time delay, $\Delta t = \hbar d\eta/dE$, where $\hbar$ is the reduced Planck constant, $\eta$ the scattering phase and $E$ the electron energy, was introduced by Eisenbud, Wigner and Smith over sixty years ago~\cite{eisenbud_1948_formal,Wigner_1955_PR,Smith_1960_PR}.
Over the past decade, attosecond spectroscopy techniques have enabled the measurement of ionization time delays in atoms and molecules  \cite{SchultzeScience2010Ne,isinger_2017_sience_photoionization,Huppert_PRL_2016Mol,Vos_2018_Science_CO,cattaneo-Nphy-2018attosecond,bustoPRL2019fano,Biswas_NatPhys2020_Mol,peschelNC2022attosecond_Ne,driver-nature-2024,loriot-Nphy-2024attosecond}, the study of the impact of molecular vibrations on electron motion \cite{cattaneo-Nphy-2018attosecond,NandiSA2020N2,borrasSA2023attosecond}, the capture of the autoionization dynamics across resonances\cite{grusonScience2016HeFano,KoturNC2016ArFano,Barreau_PRL2019NeFano,luoPRR2024Ar_Fano}, and the investigation of two-electron emission in helium and xenon atoms \cite{FeistPRL2009AttosecondCorrelation,ManssonNP2014}. Moreover, attosecond measurements have shed new light on shake-up processes in helium, where electron correlations contribute to a delay of just a few attoseconds \cite{pazourekPRL2012attosecondHeCorrelation,OssianderNP2017HeCorrelation}.   

However, the photoionization dynamics close to an Amusia-Cooper minimum remain unclear. Pioneering attosecond experiments have probed the ACM region in argon despite the significant challenge of the small cross-section \cite{Klünder_2011_PRL_Probing,Guénot_2012_PRA_Photoemission,Alexandridi_2021_PRR_Attosecond}. Experimental results show a strongly-varying {\it negative} time delay across the ACM. Theoretical models, including different levels of electron correlation, lead to vastly different results for the time delay~\cite{dahlstromJPB2014study,Dixit_2013_PRL_TDLDA,Magrakvelidze_2015_PRA_TDLDA,Pi_2018_AS_TDLDA,kheifets2020angular,Ganesan_PRA2025_Relaxation}. The random phase approximation with exchange (RPAE), which incorporates multichannel coupling and exchange-correlation  
\cite{DahlstrPRA2012Diagram,KherfetsPRA2013Time,sahaPRA2014relativistic}, gives excellent agreement with experimental results for the ionization cross section \cite{mobusPRA19933Smeasurements}, but shows a strongly-varying {\it positive} time delay. Experiment and theory give two opposite physical interpretations: According to experiment, the electron is rapidly pushed out of the potential, faster than in vacuum, while, according to the RPAE prediction, the electron is trapped and escapes only after a significant delay. As a possible explanation for this discrepancy, Alexandridi et al.  \cite{Alexandridi_2021_PRR_Attosecond} suggested that the experiment could be hampered by shake-up processes spectrally overlapping with 3$s$ ionization \cite{isinger_2017_sience_photoionization}, thus indicating the necessity to perform experiments with higher spectral resolution. Another explanation is that the theoretical description is incomplete and does not capture the entire physics of the ACM. 

This work presents a joint experimental and theoretical study of argon photoionization in the ACM spectral region. We measure the time delay for ionization in the $3s$ subshell in argon relative to that for ionization in the $3p$ subshell using the interferometric RABBIT (Reconstruction of Attosecond Beating by Interference of Two-Photon Transitions) technique \cite{PaulScience2001RABBIT,mairesseScience2003attosecond}. Our experiment's high spectral resolution allows us to disentangle direct $3s$ ionization from satellite shake-up processes, where a second electron is excited to the $4p$ or $3d$ states. 
Our results show a clear negative dip structure in the photoionization time delay across the ACM, confirming previous experiments, and are in excellent agreement with advanced theoretical calculations, including additional high-order correlation effects. We explain how these electron correlations affect the photoemission dynamics.

\paragraph*{Experimental results}

\begin{figure}[b]
    \centering
    \includegraphics[width=14cm]{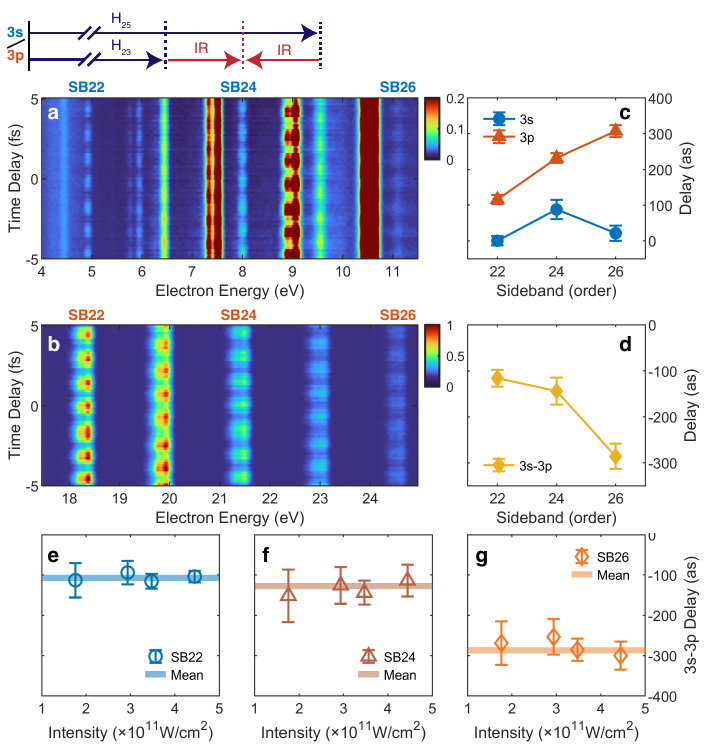}
    \caption{{\bf Attosecond interference spectroscopy.}
    Raw RABBIT spectra for {\bf a} $3s$ shell and {\bf b} $3p$ shell with photon energy range from SB22 to SB26.
    {\bf c,} Extracted phase delay of $3s$ and $3p$ channel, zeroing with SB22 of $3p$.
    {\bf d,} Relative ionization delay between $3s$ and $3p$ with probe IR intensity of $\sim$ 3.5$\times10^{11}$ W/cm$^2$.
    Extracted relative photoionization delays of {\bf e} SB22, {\bf f} SB24 and {\bf g} SB26 under different probe IR intensity.
    }
    \label{fig:Spectra}
\end{figure}

The experimental setup is described in the supplemental information \cite{SM}. Argon gas is photoionized by extreme ultraviolet (XUV) radiation in the form of odd-order harmonics and a weak fraction of the infrared (IR) laser used to generate these harmonics, delayed relative to the XUV. The photoelectron spectrum, obtained by detecting electrons in the laser and XUV polarization direction, is composed of main bands due to absorption of harmonics, and sidebands (SB$2q$) due to absorption of a harmonic of order $2q-1$ and absorption of an IR photon and to absorption of the following harmonic $2q+1$ and emission of an IR photon \cite{PaulScience2001RABBIT,mairesseScience2003attosecond}. Figures~\ref{fig:Spectra}a and \ref{fig:Spectra}b show the measured RABBIT spectrograms of the $3s$ and $3p$ ionization paths, respectively, covering the energy from 30 to 45 eV by using high-order harmonic generation (HHG) in argon. These two spectrograms show different spectral regions of the same measurement. In addition to the main and sidebands, the 3s spectrum also shows electrons due to $3p$ ionization by low-order harmonics. We limit the maximum energy of the harmonic comb to be less than 45.1 eV, thus effectively suppressing shake-up processes, such as those leaving the parent ion in the $3p^44p$ or $3p^43d$ states, which would overlap with $3s$ photoelectrons \cite{Alexandridi_2021_PRR_Attosecond}. 

The electron spectra oscillate as a function of delay with a frequency equal to $2\omega$, $\omega$ denoting the carrier frequency of the IR laser. Our interferometric measurement consists of determining the phase $\Delta\phi$ of these oscillations, representing the phase difference between the two quantum paths leading to the sideband. We show the measured group delays, $\Delta\phi/2\omega$, for the $3s$ and $3p$ channels in Fig.~\ref{fig:Spectra}c, setting to zero the $3p$ delay at SB22. $\Delta\phi/2\omega$ is the sum of the group delay of the XUV radiation $\tau_\mathrm{XUV}$ and the delay due to the ionization process or {\it ionization time delay} $\tau_{3s,3p}$. The contribution from the XUV radiation $\tau_\mathrm{XUV}$ is identical for the $3s$ and $3p$ ionization paths, allowing us to extract the difference between the ionization time delays in the $3s$ and $3p$ subshells, $\tau_{3s} - \tau_{3p}$. The results show a clear negative variation from -120 as for SB22 to -290 as for SB26 with good temporal precision (Fig.~\ref{fig:Spectra}d). 
Finally, to rule out contributions from multiphoton effects and laser-assisted Auger decay \cite{ranitovicPRL2011LaserAuger}, we varied the intensity of the IR probe pulse in the range of $1.8 - 4.6\times 10^{11}$~W/cm$^2$, as shown in Fig.~\ref{fig:Spectra}e-g. No significant changes in the ionization time delays are observed.

\begin{figure}[h]
     \centering
    \includegraphics[width=0.5\textwidth]{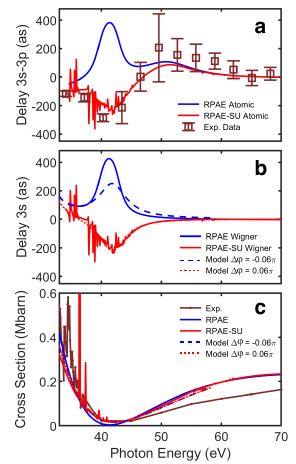}
     \caption{{\bf Photoionization time delays across the Asumia-Cooper minima.}
     {\bf a,} Measured (brown squares) and calculated relative atomic delays between ionization from Ar $3s$ and $3p.$ The calculated data are RPAE (blue line) and RPAE with selected shake-up channels (red line).
     {\bf b,} The Wigner delay for ionization of $3s$ electron obtained from RPAE (blue line) and RPAE with selected shake-up channels (red line). The Wigner delay obtained from analytical formula with $\Delta\varphi = -0.06\pi$ (blue dashed line) and $\Delta\varphi = 0.06\pi$ (red dotted line).
     {\bf c,} Cross sections calculated with RPAE (blue line) and with the same shake-up channels and relaxation as included in {\bf a}  (red line) and experimental data (brown dots line) extracted from Ref.~\cite{mobusPRA19933Smeasurements}. Fitted cross sections using analytical formula with $\Delta\varphi = -0.06\pi$ (blue dashed line) and $\Delta\varphi = 0.06\pi$ (red dotted line).     
     \label{fig:Exp_Num}}
\end{figure}

These measurements were extended using high-order harmonics generated in neon with photon energies from 45~eV to 70~eV. The difference between the ionization time delays in the $3s$ and $3p$ subshells, $\tau_{3s} - \tau_{3p}$, is plotted in Fig.~\ref{fig:Exp_Num}a over the total photon energy range, from 30 to 70 eV. The delay difference is first negative, goes through a minimum at the ACM, approaches zero around 46~eV, and finally reaches a slightly positive value near the Cooper minima of the $3p$ subshell ($\sim 51$~eV) \cite{Alexandridi_2021_PRR_Attosecond}. 

\paragraph*{Theoretical calculations} 

Fig.~\ref{fig:Exp_Num}a presents theoretical results for the difference between the ionization time delays in the $3s$ and $3p$ subshells. These calculations include two-photon interactions, simulating the RABBIT technique~\cite{dahlstromJPB2014study,Vinbladh_2019_PRA_RPAE,isinger_2017_sience_photoionization}. The results in blue are obtained with the standard RPAE approach, which includes the interplay between the $3s$ and $3p$ channels as well as ground state correlations, as described in more detail in the Supplemental Material \cite{SM}. While the agreement with experimental results is good in the 50-70 eV spectral region, the RPAE approach, giving a large positive delay difference at 42 eV ($\sim 380$~as), is not able to describe correctly the ACM spectral region. We improved our theoretical calculations by including additional electron correlation effects (see SM \cite{SM}). In particular, we found that the coupling of the $3s$-channel with {\it shake-up} channels, where a $3p$ hole is created and another $3p$ electron is excited, had a strong influence on $3s$-ionization close to the ACM. The results of this calculation, denoted RPAE-shake-up (SU), shown in red, are in excellent agreement with the experimental results. The variation of the time delay difference presents a series of resonances, characteristic of discrete excitations. 

Figure. S6 of the SM \cite{SM} shows the energy dependence of the $3s$ and $3p$ time delays calculated using the RPAE-SU and RPAE approaches. The time delay difference $\tau_{3s} - \tau_{3p}$ coincides with the $3s$ time delay across the ACM, while the positive time delay difference above 50 eV reflects the negative $3p$ time delay in the CM region. These time delays ($\tau_{3\ell})$, determined through the RABBIT technique, are not exactly the Wigner time delays ($\tau^\mathrm{W}_{3\ell}$), which describe the electron wavepacket emitted after the absorption of a single photon \cite{Wigner_1955_PR,eisenbud_1948_formal,Smith_1960_PR}. The two delays are related by $\tau_{3\ell}=\tau^\mathrm{W}_{3\ell}+\tau^\mathrm{cc}_{3\ell}$, where $\tau^\mathrm{cc}_{3\ell}$ represent a correction due to continuum-continuum transitions involved in the RABBIT technique \cite{DahlstrCP2013Review,Bustochapter}. These contributions, however, are significant only at the $3s$ ionization threshold and do not lead to a strong variation of the delay across the ACM, see Fig. S6 of the SM \cite{SM}.

In Fig.~\ref{fig:Exp_Num}b-c, we focus on single photoionization in the $3s$ subshell. The $3s$ Wigner delays calculated using the RPAE (blue) and the RPAE-SU (red) approaches are plotted as a function of energy in Fig.~\ref{fig:Exp_Num}b. The RPAE-SU calculation predicts opposite Wigner delay variation across the ACM \cite{DahlstrPRA2012Diagram,KherfetsPRA2013Time}. 
In contrast, the cross-section in Fig.~\ref{fig:Exp_Num}c is rather insensitive to the inclusion of shake-up channel interactions, apart from the resonances mentioned above, known from experiments~\cite{mobusPRA19933Smeasurements} but absent in RPAE by construction.
Electron correlations influence cross sections (amplitudes) and delays (phases) very differently, underlining the new and complementary information obtained from attosecond delay measurement.

\begin{figure}[h]
     \centering
    \includegraphics[width=0.6\textwidth]{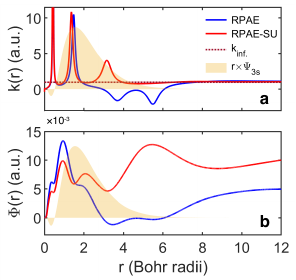}
     \caption{{\bf Photoelectron behavior within the core.}
     This figure shows {\bf a} the local wavevector $k(r)$ and
     {\bf b} the probability density flux, $\Phi (r)$, after ionization from the $3s$-orbital. 
     The result with RPAE is shown in blue, and that with  shake-up configurations added in red. In both cases the depicted situation is after the absorption of a $42$~eV photon, i.e. the photon energy is tuned to the ACM region. The $3s$ source term in the equation for the outgoing wave function is given by the orange area.
     An outward rate is positive, and an inward rate is negative. The red dotted lines gives the asymptotic $k$, i. e. when the electron has left the atomic potential.
    \label{fig:K_Flux}}
\end{figure}

To get a more intuitive picture, we study the behavior of the numerically constructed outgoing wave function $\rho(r)$ for a monochromatic harmonic field \cite{dahlstromJPB2014study}. We chose a photon energy equal to 42 eV, i.e. at the ACM. We present first the local wave vector, $k(r)=d\arg[\rho(r)]/dr$ in Fig.~\ref{fig:K_Flux}a for the two cases, RPAE (blue) and RPAE-SU (red). The $3s$ source term in the equation for the outgoing wave function is shown for reference as an orange-shaded area \cite{SM}. 
The red curve always shows a positive local wave vector, which is further mainly above the asymptotic value (red dotted line), where the latter is indicating that the electron is moving faster than the free electron.
In contrast, the blue curve becomes negative for $r$ between 3 and 6 Bohr radii. In this case, the electron is delayed. Finally, we compute the flux as a function of the radial distance, $r$:
\begin{equation}
\Phi(r) = \left.\frac{i\hbar}{2m_e}
\left[
\rho^*(s)\frac{d\rho}{ds}-\rho(s)\frac{d\rho^*}{ds}
\right]\right|_{s=r}, 
\label{eq:flux}
\end{equation}
The results, see Fig.~\ref{fig:K_Flux}b, show a striking difference between the two calculations. In the RPAE case, the flux becomes negative between 3 and 6 Bohr radii. This corresponds to an inward particle flow that may be caused by potential barriers, resulting in an increased outward flow in other channels. Thus, the electron is temporarily trapped. In contrast, the inclusion of shake-up correlations results in a positive rate everywhere, leading asymptotically to a slightly higher cross section than in the RPAE case (Fig.~\ref{fig:Exp_Num}c).

\paragraph*{Wavepacket analysis}
Our theoretical results show that correlation effects can dramatically change the sign of the ACM Wigner delay without significantly changing the cross section. Here, we propose a simple model to illustrate this observation based on the interpretation of the ACM as an interference effect between two dipoles, $z_0$ corresponding to the uncorrelated photoionization from the $3s$ orbital, and $\delta z_\pm$ describing the effects of electron correlation \cite{Amusia_1972_PL_s_subshell}. The resulting dipole $z_\pm$ can be expressed as a function of the photon energy $\omega$ as
\begin{equation}\label{eq:analytical}
z_\pm(\omega)=z_0(\omega) + \delta z_\pm(\omega) = z_0(\omega) \left[1-\kappa e^{\pm i\Delta\varphi} \mathrm{atan}\left(\frac{\omega-\epsilon_z}{\Delta \epsilon_z}\right) \right],
\end{equation}
The relative coupling between the two dipoles is described by a complex-valued constant $\kappa e^{\pm i\Delta\varphi}$ ($\Delta \varphi$ and $\kappa$ being positive) multiplied by an arc-tangent function centred at $\epsilon_z$ and with a width $\Delta\epsilon_z$ to represent the variation of a transition amplitude close to the Cooper minimum. Clearly, the cross section is insensitive to the sign of the correlation phase, since $|z_+(\omega)|^2=|z_-(\omega)|^2$.
The model parameters are fitted to reproduce the experimental and theoretical results for the cross section and the Wigner delay. Excellent agreement is obtained as shown by the red dotted line in Fig.~\ref{fig:Exp_Num}b and \ref{fig:Exp_Num}c. The phase used in this model is remarkably small, $\Delta\varphi=0.06\pi$. Changing its sign leads to the dashed blue curve, which is qualitatively similar to the RPAE result. In this way, the analytical model explains how a modest change in the phase ($0.12\pi$) can dramatically alter the Wigner delay without {\it any} change in the cross section. 
While it was already known that modest changes in the description of correlation effects can change the sign of the delay \cite{dahlstromJPB2014study}, and 
that this phase ambiguity can be related to topology 
\cite{jiNJP2024relation}, our study is the first systematic expansion of the ACM beyond the RPAE. 
Further details about the analytical model are given in \cite{SM}.

\begin{figure}[h]
    \centering
    \includegraphics[width=14cm]{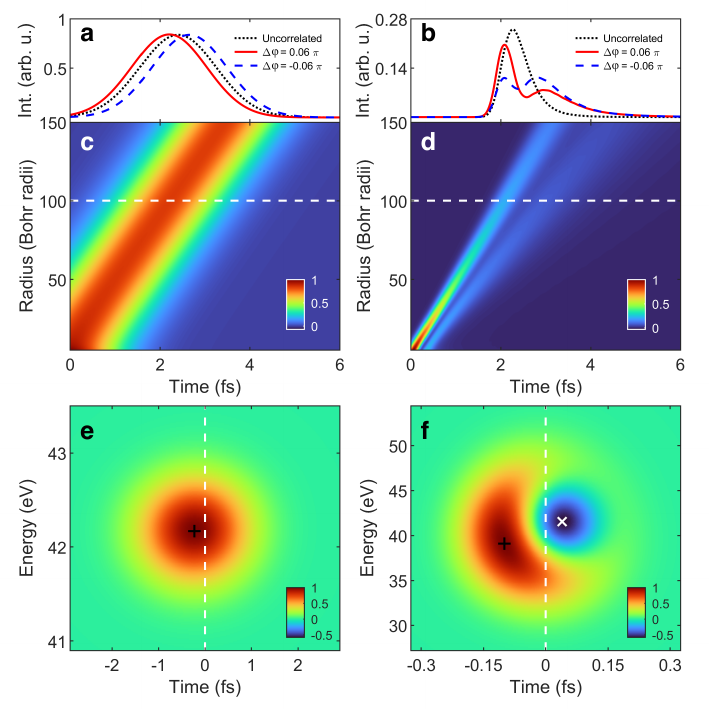}
    \caption{{\bf Wave packets and Wigner Distributions.} 
    {\bf a} and {\bf b}, The wavepacket distributions obtained with a 2~fs pulse and a 200~as XUV light pulse when propagated to 100 a.u.. {\bf c} and {\bf d}, The time dependent fluxes which leading negative Wigner delay interacting with a 2~fs pulse and a 200~as XUV light pulse. {\bf e} and {\bf f}, The corresponding Wigner distributions including correlation that leads to a negative Wigner delay interacting with a 2~fs and a 200~as XUV light pulse.
    }
    \label{fig:time_domain}
\end{figure}

This simple model allows us to study the dynamics of the electron wave packet created by the absorption of ultrashort pulses in the ACM region. Fig.~\ref{fig:time_domain}a, c, e present results obtained for a pulse of $\tau=2\,$ fs, while Fig.~\ref{fig:time_domain}b, d, f those for a pulse of $\tau=200\,$as. Fig.~\ref{fig:time_domain}a shows the time-dependent photoelectron flux $\Phi(r,t)$ at $r=100$ Bohr radii for $\pm \Delta\varphi = \pm 0.06\pi$ and for $\kappa =0$, i.e. the uncorrelated case. Compared to the latter case, the electron wave packet is advanced ($0.06\pi$) or delayed ($-0.06\pi$). Fig.~\ref{fig:time_domain}b presents the same quantities calculated for a 200 as pulse. The electron wave packets are created with broad energy components that span the entire ACM spectral structure and unfold a more complex behavior in time. Unlike the uncorrelated wave packet, the correlated wave packets have a double-peaked structure with a minimum in between, consisting of early and late electrons compared to the uncorrelated electrons. 
Fig.~\ref{fig:time_domain}c and \ref{fig:time_domain}d show the flux of the outgoing wave function as a function of time and radius. In the case of a short pulse, the wave packet with a positive (negative) Wigner delay peak shows a stronger late (early) peak. It is a time domain manifestation of the Amusia interference effect, which can develop as the photoelectron is dispersed in time and space due to free propagation. 

The time-frequency representation of electron wave packets has been a useful tool in recent years to understand details of the photoemission process \cite{4dxenon_Zhong2020,decoherence_Busto2022}. Here, we analyze our wave packets using the {\it Wigner transform}
\begin{equation}
{\cal Z}(t,\omega)=\frac{1}{\pi}\int d\omega' Z^*(\omega+\omega') Z(\omega-\omega')e^{i2\omega't},
\end{equation}
where ${\cal Z}(t,\omega)$ is a quasi-probability distribution in time and frequency. The effective dipole $Z(\omega)=z_+(\omega)g_{\omega_0,\tau}(\omega)$ is the product of the analytical dipole element, $z_+(\omega)$ [see Eq.~(\ref{eq:analytical})] and a spectral-filter function, $g_{\omega_0,\tau}(\omega)=\exp[-\frac{\tau^2}{8\ln(2)}(\omega-\omega_0)^2]$, which selects the spectral bandwidth around the ACM corresponding to a pulse with duration $\tau$. The quasi-probability distribution for a 2 fs pulse (Fig.~\ref{fig:time_domain}e) exhibits a Gaussian distribution that is shifted in time by $-230\,$as (see $+$ symbol) which is in excellent agreement with the obtained $3s$ Wigner delay (Fig.~\ref{fig:Exp_Num}b). In contrast, for a 200 as pulse (Fig.~\ref{fig:time_domain}f), the quasiprobability distribution has a maximum located at $-100\,$as, indicated by the $+$ symbol. The interference structure represented by a negative quasi-probability with minimum at $40\,$as, marked by $\times$, is due to electron correlation, and results in the minima in the spectral domain (Fig.~\ref{fig:Exp_Num}b-c) and in the time-space domain (Fig. \ref{fig:time_domain}c-d). 
This wavepacket analysis reveals that the influence of electronic correlation leads to time-space and time-frequency interferences of the electron wavepacket, provided a sufficiently short pulse is used. 

\paragraph*{Conclusion}

In summary, we have measured with high spectral precision the difference in attosecond ionization time delays between the $3s$ and $3p$ subshells across the Amusia-Cooper minimum of argon. 
We have extended the RPAE theoretical approach, which takes into account coupling between $3s$ and $3p$ ionization, to include also couplings to shake-up channels. A wavepacket analysis 
illustrates the dynamics of the interaction close to the Amusia-Cooper minimum. 

Our results reveal that electron correlation leads to an advance of the electron emitted from the $3s$ subshell relative to a free electron of $\sim 240$~as. High-order electron-correlation effects, mediated by shake-up configurations, strongly affect the temporal dynamics of the escaping electron.
While traditional cross-section measurements are essentially insensitive to these correlations, we demonstrate that the attosecond ionization time delay is a fundamental physical quantity that provides unprecedented insight into electron correlation. Our work lays the foundations for studying the dynamics in multi-electronic systems with broader implications for tracing electron interactions of inner orbital vacancies in atoms and molecules.


\bibliography{scibib}

\bibliographystyle{Science}

\section*{Acknowledgments}

\paragraph*{Funding:}
S.~L. and D.~D. acknowledge support from National Natural Science Foundation of China (Grants No.12450402, No.12134005, and No.11627807). A.~L., M.~G. and E.~L. acknowledge support from the Swedish Research Council (Grant Nos. 2020-05200, 2023-04603, 2024-05934, 2020-03315, 2024-04115), and from the Knut and Alice Wallenberg Foundation. A.~L. and D.~B. acknowledges support from the Knut and Alice Wallenberg Foundation through the Wallenberg Centre for Quantum Technology (WACQT). A.~L. acknowledges support from the European Research Council (advanced grant QPAP, Grant No. 884900). P.~V.~D. acknowledge support by the DFG Project No. 492619011 (DE 2366/6-2).

\paragraph*{Author contributions:}
M.L., H.W., J.L, W.W., J.C., X.H., X.S., X.Z., and S.L. performed the experiment and collected the data.
E.L. developed the RPAE-SU approach and provided the numerical results.
R.T. and J.M.D. developed the analytical model.
P.V.D. contributed by calculations that helped to formulate and initiate the theoretical study at hand.
M.L. and S.L. proposed the experimental methodology.
A.L., E.L., J.M.D., D.D. and S.L. led the project.
M.L., R.T., R.W., D.B., M.G., A.L., J.M.D., E.V. and S.L. interpreted the data and contributed to the preparation of the manuscript.
All authors discussed and approved the results and contributed to the revision of the manuscript.

\paragraph*{Competing interests:}
There are no competing interests to declare.

\paragraph*{Data and materials availability:}
All data are available in the main text or the supplementary materials.

\section*{Supplementary materials}
Materials and Methods\\
Figs. S1 to S6\\
References \textit{(1-14)}


\clearpage

\end{document}